\documentclass[%
 reprint,
 amsmath,amssymb,
 aps,
]{revtex4-2}

\usepackage{graphicx}
\usepackage{dcolumn}
\usepackage{bm}

\usepackage{color}     
\usepackage{amsmath}

\begin{document}

\preprint{APS/123-QED}

\title{Nuclear $\beta$ spectrum from projected shell model (I): allowed one-to-one transition}

\author{Fan Gao}%
\affiliation{School of Physical Science and Technology, Southwest University, Chongqing 400715, China}%

\author{Zi-Rui Chen}%
\affiliation{School of Physical Science and Technology, Southwest University, Chongqing 400715, China}%

\author{Long-Jun Wang}
\email{longjun@swu.edu.cn}
\affiliation{School of Physical Science and Technology, Southwest University, Chongqing 400715, China} 

\date{\today}

\begin{abstract}
  Nuclear $\beta$ spectrum and the corresponding (anti-)neutrino spectrum play important roles in many aspects of nuclear astrophysics, particle physics, nuclear industry and nuclear data. In this work we propose a projected shell model (PSM) to calculate the level energies as well as the reduced one-body transition density (ROBTD) by the Pfaffian algorithm for nuclear $\beta$ decays. The calculated level energies and ROBTD are inputed to the Beta Spectrum Generator (BSG) code to study the high precision $\beta$ spectrum of allowed one-to-one transitions. When experimental level energies are adopted, the calculated $\beta$ spectrum by ROBTD of the PSM deviates from the one by the extreme simple particle evaluation of the BSG by up to $10\%$, reflecting the importance of nuclear many-body correlations. When calculated level energies are adopted, the calculated $\beta$ spectrum shows sensitive dependence on the reliability of calculated level energies. The developed method for ROBTD by the PSM will also be useful for study of the first-forbidden transitions, the isovector spin monopole resonance etc. in a straightforward way. 
\end{abstract}

\maketitle


\section{\label{sec:intro}Introduction}

Nuclear $\beta$ decay plays crucial roles in many frontiers in nuclear astrophysics and nuclear physics. For nuclear astrophysics, the understanding of many astrophysical problems such as the cooling of neutron stars \cite{schatz2014nature, LJWang_2021_PRL}, the origin of heavy elements by rapid neutron and proton capture processes etc. \cite{langanke_RMP, r_process_RMP_2021, rp_process_Schatz_1998} rely heavily on effective $\beta$-decay rates of finite nuclei in stellar environments with high temperature and high density. 

The nuclear $\beta$ spectrum, and the neutrino spectrum accordingly, i.e., the probability of the electron and anti-neutrino emission (positron and neutrino emission) as a function of the corresponding kinetic energy for $\beta^-$ decay ($\beta^+$ decay), are important for nuclear astrophysics, particle physics, nuclear industry and nuclear data \cite{Mougeot_PRC_2015, Hayen_RMP_2018, Hayen_CPC_code_2019, Kumar_2nd_forbidden_2020}. The integrated form of the $\beta$ spectrum turns out to be the most precise way to determine the up-down quark element of the Cabibbo-Kobayashi-Maskawa matrix $V_{ud}$ \cite{Hardy_PRC_2015}. The $\beta$ spectrum is also needed in nuclear medicine for radiotherapy and dosimetry, and in the nuclear industry for calculations of residual power or post-irradiation fuel management \cite{Bardies_1994, Mougeot_PRC_2015}. The neutrino spectra of reactors provide one of the techniques for monitoring nuclear facilities, and the reactor anti-neutrino anomaly is still a puzzle to be solved for the community \cite{Mention_PRD_2011, Fallot_PRL_2012, Haag_PRL_2014, Zakari_PRL_2015, An_PRL_2016, Choi_PRL_2016, Rasco_PRL_2016, An_PRL_2017, Fijia_PRL_2017, Sonzogni_PRL_2017, Hayes_PRL_2018, Guadilla_PRL_2019, Estienne_PRL_2019, Kharusi_PRL_2020, An_PRL_2022, Almaz_PRL_2022}. 

\begin{figure}[htbp]
\begin{center}
  \includegraphics[width=0.48\textwidth]{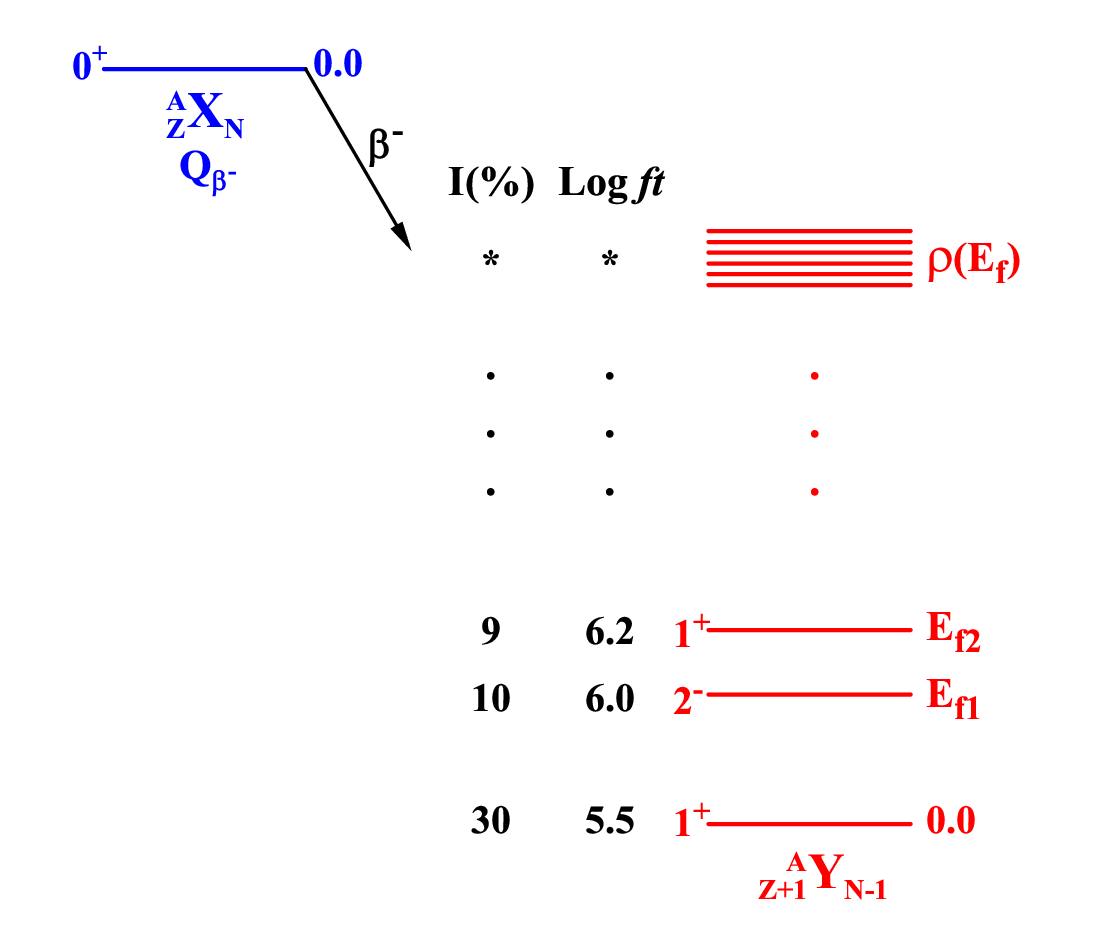}
  \caption{\label{fig:fig_1} (Color online) A schematic diagram for the $\beta$ decay scheme of some even-even nucleus $^{A}$X.} 
\end{center}
\end{figure}

In practical investigations and applications, $\beta$ spectra and neutrino spectra of many nuclei, including some exotic ones that cannot be touched experimentally so far, are indispensable so that both experimental measurements and theoretical calculations are expected. Accordingly, some aspects need to be noted for the study of nuclear $\beta$ and neutrino spectra, following the schematic diagram of the $\beta$ decay scheme as shown in Fig. \ref{fig:fig_1}. (i) One-to-one transition spectrum and the total spectrum. As seen from Fig. \ref{fig:fig_1}, the initial state (usually the ground state while sometimes an isomer) of some parent nucleus can $\beta$ decay to each state of the corresponding daughter nucleus within the $Q_{\beta}$ value, where each one-to-one (individual) transition has the corresponding feeding intensity $I(\%)$ and comparative half-life Log$ft$. Each individual transition has corresponding one-to-one transition $\beta$ spectrum and the total $\beta$ spectrum can be obtained theoretically by summing over all related normalized one-to-one transition $\beta$ spectrum weighted by corresponding $I(\%)$. This indicates that reliable calculation of the total $\beta$ spectrum depends on reasonable description of both the one-to-one transition $\beta$ spectra and the $I(\%)$. (ii) Allowed and forbidden transitions. Not only allowed but also forbidden transitions can provide sizable contributions to the total $\beta$ spectrum. As seen from the schematic diagram in Fig. \ref{fig:fig_1}, the transition to the $1^+$ state ($2^-$ state) with excitation energy $E_{f2}$ ($E_{f1}$) is allowed (first-forbidden), both transitions may have similar Log$ft$ and $I(\%)$ so that they may provide similar contribution to the total spectrum. Actually, there are many nuclei whose $I(\%)$ is dominated by a single forbidden transition (such as $^{92}$Rb, $^{89}$Sr, $^{96}$Y, $^{137}$Xe, $^{142}$Cs etc.), and thus the total spectrum is determined by reliable description of the forbidden transition theoretically. (iii) Contribution of highly-excited states. Nuclear levels are usually discrete for low-lying states, and become denser and denser with excitation energy so that the level density $\rho(E)$ is large for highly-excited states (especially of odd-mass and odd-odd nuclei). For $\beta$ decay with large $Q_{\beta}$, the contribution from highly-excited states with large $\rho(E)$ should also be considered accurately and precisely in the total $\beta$ spectrum. Experimentally, the $I(\%)$ of highly-excited states cannot be obtained accurately by the traditional method for $\gamma$-ray spectroscopy measurements with the high purity germanium (HPGe) arrays which have high energy resolution but low efficiency. This is referred to as the Pandemonium effect \cite{Pandemonium_Hardy_PLB_1977}. The problem can be overcome by adopting inorganic scintillators such as NaI that have very high efficiency but reduced energy resolution, corresponding to the modern total absorption $\gamma$-ray spectroscopy (TAGS) technique \cite{Guadilla_PRL_2019, Shuai_PRC_2022}. Theoretically, large model space and configuration space are crucial for reliably modeling the $I(\%)$ of highly-excited states. 

It can then be seen that reliable calculation of the total $\beta$ and neutrino spectra requires nuclear-structure models to treat reasonably both allowed and forbidden transitions, to have large model and configuration spaces for description of highly-excited states, to work in the laboratory frame with good angular momentum and parity as allowed and forbidden transitions have strong and different selection rules, and to treat nuclei from light to superheavy ones. The projected shell model (PSM) \cite{PSM_review, Sun_1996_Phys_Rep}, when updated with extended large configuration space \cite{LJWang_2014_PRC_Rapid, LJWang_2016_PRC}, can serve as one of the candidate models for nuclear $\beta$ and neutrino spectra. The PSM has been applied successfully on studies of nuclear levels \cite{LJWang_2019_JPG, LJWang_PLB_2020_chaos} and weak-interaction processes \cite{LJWang_2018_PRC_GT, LJWang_PLB_2020_ec, LJWang_2021_PRL, LJWang_2021_PRC_93Nb} recently, with the help of some modern algorithms \cite{ZRChen_2022_PRC, BLWang_2022_PRC}. The exact angular-momentum-projection technique that is adopted by the PSM is also crucial for study of rare nuclear weak decays by other models \cite{LJWang_current_2018_Rapid, Yao_Wang_0vbb_PRC_2018}. We aim at developing the PSM further for description of nuclear total $\beta$ and neutrino spectra step by step. In this work, we develop the PSM for calculation of the reduced one-body transition density (ROBTD) by the Pfaffian algorithm for nuclear $\beta$ decays. On one hand, the ROBTD and level energy are indispensable nuclear inputs for the modern Beta Spectrum Generator (BSG) code for highly precision theoretical description of the allowed one-to-one $\beta$ spectrum \cite{Hayen_CPC_code_2019}. On the other hand, the ROBTD is also one of the two indispensable ingredients for description of forbidden transitions. The latter has been accomplished and will be published elsewhere soon. 

The paper is organized as follows. In Sec. \ref{sec:BSG} we introduce briefly the basic framework for $\beta$ spectrum of allowed one-to-one transition and the BSG, in Sec. \ref{sec:PSM} the PSM for ROBTD is presented. The $\beta$ spectra for $^{162}$Gd and $^{103}$Ru are discussed in Sec. \ref{sec:result} and we finally summarize our work in Sec. \ref{sec:sum}.

\section{\label{sec:BSG}General form of $\beta$ spectrum for allowed transition }

Following Refs. \cite{Hayen_RMP_2018, Hayen_CPC_code_2019}, after introducing several corrections including atomic effects such as the screening and exchange processes, the analytical $\beta$ spectrum of allowed one-to-one transitions with very high precision can be written as,
\begin{eqnarray} \label{eq.spectrum}
  \frac{dN(\omega)}{d\omega} &=& \frac{G_V^2 V_{ud}^2}{2\pi^3} F(Z, \omega) p\omega(Q_{if} - \omega)^2  \nonumber \\
  & \times &  C(Z, \omega) R(Q_{if}, \omega) X(Z, \omega) r(Z, \omega) \nonumber \\
  & \times &  U(Z, \omega) D(Z, \omega, \beta_2) R_N(Q_{if}, \omega) Q(Z, \omega) S(Z, \omega) \nonumber \\
\end{eqnarray}
where $\omega$ ($p=\sqrt{\omega^2-1}$) is the total energy (the momentum) of the electron in unit of $m_e c^2$ ($m_e c$), $Z$ is the proton number of the daughter nucleus and $G_V$ the vector coupling strength. $Q_{if}$ labels the available energy for leptons of the transition, i.e., 
\begin{eqnarray} \label{eq.Qif}
  Q_{if} = (\Delta M_{pd} + E_i - E_f) / m_e c^2
\end{eqnarray}
where $\Delta M_{pd}$ denotes the nuclear-mass difference of parent and daughter nuclei and $E_i$ ($E_f$) is the excitation energy of initial (final) state of parent (daughter) nucleus. It is noted that $E_i$ is zero (has finite value) for the ground state (some isomer) of the decaying parent nucleus. 

In Eq. (\ref{eq.spectrum}) $F(Z, \omega)$ is the Fermi function reflecting the Coulomb distortion of electron radial wave function by the nuclear charge, where nuclear charge distribution is adopted as a uniformly charged sphere. $p\omega(Q_{if} - \omega)^2$ is the phase space factor. These two terms dominate the $\beta$ spectrum shape and are then the most crucial parts for $\beta$ spectrum. It can be seen from Eqs. (\ref{eq.spectrum}, \ref{eq.Qif}) that nuclear level energies ($E_i, E_f$) are important for $\beta$ spectrum as they affects the phase space factor. The following four terms are the main corrections to the $\beta$ spectrum, in which $C(Z, \omega)$ is the shape factor where radial wave function behavior and all nuclear-structure-sensitive information are included. The latter is concerned with four nuclear form factors relevant to allowed $\beta$ decay (see Table \ref{tab:table1}), which, following some approximations such as the impulse approximation, can be transformed into nuclear matrix elements of corresponding one-body operators, i.e., \cite{Hayen_RMP_2018, Hayen_CPC_code_2019}
\begin{eqnarray} \label{eq.RME}
  & & \big\langle \Psi^{n_f}_{J_f} \big\| \hat O_\lambda \hat\tau^{\pm} \big\| \Psi^{n_i}_{J_i} \big\rangle \nonumber \\
  &=& \hat\lambda^{-1} \sum_{\mu\nu} \langle \mu \| \hat O_\lambda \hat\tau^{\pm} \| \nu \rangle 
      \big\langle \Psi^{n_f}_{J_f} \big\| \left[ \hat c^\dag_\mu \otimes \tilde{\hat c}_\nu \right]^\lambda \big\| \Psi^{n_i}_{J_i} \big\rangle
\end{eqnarray}
where $\hat\lambda \equiv \sqrt{2\lambda+1}$ with $\lambda$ being the rank of operators, $|\Psi^{n}_{J}\rangle$ labels the nuclear many-body wave function for the $n$-th eigen-state of angular momentum $J$, $\hat\tau^\pm$ is the isospin ladder operator for $\beta^\pm$ decay so that $\mu$ and $\nu$ are single-particle proton (neutron) and neutron (proton) states for $\beta^-$ ($\beta^+$) decay. Here we adopt the spherical harmonic oscillator basis so that $|\mu\rangle \equiv |n_\mu, l_\mu, j_\mu \rangle$ and $\hat c^\dag, \tilde{\hat c}$ are the corresponding single-particle creation and annihilation operators in the form of irreducible spherical tensor. The last term in Eq. (\ref{eq.RME}) is the ROBTD which characterize the many-nucleon properties of the initial ($i$) and final ($f$) nuclear states \cite{Suhonen_book}. 

\begin{table}[t] 
  \caption{\label{tab:table1} The four form factors for allowed Gamow-Teller decay. The first column shows the notation by Behrens and B\"uhring (BB) \cite{BB_book_1982} and the second shows the evaluation of the form factor in the impulse approximation in cartesian coordinates, where $M_N$ and $R$ are the nucleon mass and nuclear radius, $\bm\sigma$ the Pauli matrix, $\beta=-\gamma_0$ and $\bm\alpha=\gamma_5\bm\sigma$ with the Dirac $\gamma$-matrices, $g_A, g_M, g_P$ the coupling constant for axial, weak magnetism and induced pseudoscalar currents respectively, and the summation over nucleons and the isospin ladder operators are not shown for convenience. Refer to Refs. \cite{Hayen_RMP_2018, Hayen_CPC_code_2019} for details. }
  \begin{ruledtabular}
  \begin{tabular}{cc}
    \textrm{Form Factor}              & \textrm{Matrix element}           \\
    \textrm{(BB)}  & \textrm{(Impulse Approximation)}  \\ \hline
    $^A$F$^{(0)}_{101}$   & $\mp g_A \int \bm\sigma$     \\
    $^A$F$^{(0)}_{121}$   & $\mp g_A \frac{3}{\sqrt 2} \int \frac{(\bm\sigma\cdot\bm r)\bm r - \frac{1}{3}\bm\sigma\cdot r^2}{R^2} \mp g_P \frac{2M_N}{R}5\sqrt{2} \int \beta\gamma_5 \frac{i\bm r}{R}$     \\
    $^V$F$^{(0)}_{111}$   & $- g_V \sqrt{\frac{3}{2}} \int \frac{\bm\alpha\times\bm r}{R} - (g_M-g_V) \frac{2M_N}{R} \sqrt{6} \int \bm\sigma$ \\
    $^A$F$^{(0)}_{110}$   & $\mp g_A \sqrt{3} \int \gamma_5 \frac{i\bm r}{R}$  \\
  \end{tabular}
  \end{ruledtabular}
\end{table}

The $R(Q_{if}, \omega)$ term in Eq. (\ref{eq.spectrum}) is the radiative correction corresponding to higher-order loop corrections for the interaction of the $\beta$ particle and nucleus beyond the Fermi function \cite{Hayen_RMP_2018, Hayen_CPC_code_2019}. The $X(Z, \omega)$ term is the correction for atomic exchange effect corresponding to the possibility of the $\beta$ particle to be emitted into a bound state of the daughter atom by expelling the bound electron into the continuum. The $r(Z, \omega)$ term is the atomic mismatch correction that takes into account discrete processes such as shake-off and shake-up \cite{Hayen_RMP_2018, Hayen_CPC_code_2019}. The last five terms in Eq. (\ref{eq.spectrum}) are minor corrections that account for high-precision details of the $\beta$ spectrum, where $U(Z, \omega)$ is for the diffused nuclear charge distribution near nuclear surface, $D(Z, \omega, \beta_2)$ for the small effect of nuclear deformation for charge distribution, $R_N(Q_{if}, \omega)$ for the effect of the recoil of a nucleus after $\beta$ decay, $ Q(Z, \omega)$ for the recoil Coulomb correction and $S(Z, \omega)$ for the atomic screening correction that the $\beta$ particle feels effective nuclear charge due to the screening of the orbiting electrons of the atom. 

The analytical expressions of all the above corrections to allowed $\beta$ spectrum have been studied in Ref. \cite{Hayen_RMP_2018} in details, and a modern BSG code is developed in Ref. \cite{Hayen_CPC_code_2019} to take into account all these corrections numerically for high-precision allowed $\beta$ spectrum shapes, provided that some important nuclear-structure informations such as the level energies and the ROBTD should be inputed. In this work we install, study, test and employ the BSG code successfully and develop the PSM model for nuclear level energies and ROBTD successfully, then combine them for the study of allowed one-to-one transition $\beta$ spectrum shapes.

\section{\label{sec:PSM}Projected shell model for levels and ROBTD}

It is expected but not easy to develop a practical nuclear-structure model that can treat reasonably both allowed and forbidden transitions, having large model and configuration spaces for description of highly-excited states, working in the laboratory frame with good angular momentum and parity for nuclear states as allowed and forbidden transitions have strong and different selection rules, and can treat even-even, odd-mass and odd-odd nuclei from light to superheavy ones. We aim to accomplish this goal by further developing the PSM model extensively. 

Due to the complexity of nuclear many-body problem, one of the practical solutions is to reduce the many-body to effective one-body problem by mean-field calculations in the intrinsic frame, then consider residual many-body correlations by beyond-mean-field methods and get the nuclear many-body
wave functions in the laboratory frame. Following this philosophy, the PSM \cite{PSM_review} stars from Nilsson+BCS mean-field calculations (which can be generalized to elaborated ones for low-lying states \cite{LJWang_current_2018_Rapid, Yao_Wang_0vbb_PRC_2018}) and then adopts configuration mixing and angular-momentum projection (AMP) for beyond mean-field part. The configuration spaces of the PSM are then,
\begin{align} \label{eq.config}
  \textrm{ee}: \big\{ & |\Phi \rangle, 
               \hat{a}^\dag_{\nu_i} \hat{a}^\dag_{\nu_j} |\Phi \rangle,
               \hat{a}^\dag_{\pi_i} \hat{a}^\dag_{\pi_j} |\Phi \rangle,
               \hat{a}^\dag_{\nu_i} \hat{a}^\dag_{\nu_j} \hat{a}^\dag_{\pi_k} \hat{a}^\dag_{\pi_l} |\Phi \rangle, \nonumber\\ 
             & \hat{a}^\dag_{\nu_i} \hat{a}^\dag_{\nu_j} \hat{a}^\dag_{\nu_k} \hat{a}^\dag_{\nu_l} |\Phi \rangle, 
               \hat{a}^\dag_{\pi_i} \hat{a}^\dag_{\pi_j} \hat{a}^\dag_{\pi_k} \hat{a}^\dag_{\pi_l} |\Phi \rangle,  \cdots \big\} \nonumber \\
  \textrm{o}\nu: \big\{ & \hat{a}^\dag_{\nu_i} |\Phi \rangle,
                 \hat{a}^\dag_{\nu_i} \hat{a}^\dag_{\nu_j} \hat{a}^\dag_{\nu_k} |\Phi \rangle, 
                 \hat{a}^\dag_{\nu_i} \hat{a}^\dag_{\pi_j} \hat{a}^\dag_{\pi_k} |\Phi \rangle,   \nonumber \\
               & \hat{a}^\dag_{\nu_i} \hat{a}^\dag_{\nu_j} \hat{a}^\dag_{\nu_k} \hat{a}^\dag_{\pi_l} \hat{a}^\dag_{\pi_m} |\Phi \rangle,
                 \cdots \big\} \nonumber \\
  \textrm{o}\pi: \big\{ & \hat{a}^\dag_{\pi_i} |\Phi \rangle,
                 \hat{a}^\dag_{\pi_i} \hat{a}^\dag_{\pi_j} \hat{a}^\dag_{\pi_k} |\Phi \rangle, 
                 \hat{a}^\dag_{\pi_i} \hat{a}^\dag_{\nu_j} \hat{a}^\dag_{\nu_k} |\Phi \rangle,   \nonumber \\
               & \hat{a}^\dag_{\pi_i} \hat{a}^\dag_{\pi_j} \hat{a}^\dag_{\pi_k} \hat{a}^\dag_{\nu_l} \hat{a}^\dag_{\nu_m} |\Phi \rangle,
                 \cdots \big\} \nonumber \\
  \textrm{oo}: \big\{ & \hat{a}^\dag_{\nu_i} \hat{a}^\dag_{\pi_j}|\Phi \rangle, 
               \hat{a}^\dag_{\nu_i} \hat{a}^\dag_{\nu_j} \hat{a}^\dag_{\nu_k} \hat{a}^\dag_{\pi_l} |\Phi \rangle,
               \hat{a}^\dag_{\nu_i} \hat{a}^\dag_{\pi_j} \hat{a}^\dag_{\pi_k} \hat{a}^\dag_{\pi_l} |\Phi \rangle, \nonumber\\ 
             & \hat{a}^\dag_{\nu_i} \hat{a}^\dag_{\nu_j} \hat{a}^\dag_{\nu_k}  
               \hat{a}^\dag_{\pi_l} \hat{a}^\dag_{\pi_m} \hat{a}^\dag_{\pi_n} |\Phi \rangle,  \cdots \big\} 
\end{align}
for even-even (ee), odd-neutron (o$\nu$), odd-proton (o$\pi$) and odd-odd (oo) nuclei respectively, where $|\Phi \rangle$ is the quasiparticle (qp) vacuum with associated intrinsic deformation and $\hat{a}^\dag_\nu (\hat{a}^\dag_\pi)$ labels neutron (proton) qp creation operator. The large configuration spaces in Eq. (\ref{eq.config}) are developed in Refs. \cite{LJWang_2014_PRC_Rapid, LJWang_2016_PRC, LJWang_2018_PRC_GT} recently and three or more major harmonic-oscillator shells can be adopted for model space. 

The many-body configurations in Eq. (\ref{eq.config}) cannot be adopted as the many-body basis as many symmetries are broken in the mean-field calculations in the intrinsic frame. The projection technique can restore broken symmetries in the intrinsic frame and transform the configurations to projected basis in the laboratory frame \cite{Ring_many_body_book}. For example, the broken rotational symmetry in deformed intrinsic mean fields can be restored by the AMP operator,
\begin{eqnarray} \label{AMP_operator}
    \hat{P}^{J}_{MK} = \frac{2J + 1}{8\pi^2} \int d\Omega D^{J\ast}_{MK} (\Omega) \hat{R} (\Omega) ,
\end{eqnarray}
where $\hat{R}$ and $D_{MK}^{J}$ (with Euler angle $\Omega$) \cite{Angular_momentum_Varsha_book_1988} are the rotation operator and Wigner $D$-function \cite{BLWang_2022_PRC} respectively. The nuclear many-body wave function in the laboratory system can then be expressed in the projected basis, 
\begin{eqnarray} \label{eq.wave_function}
  | \Psi^{n}_{JM} \rangle = \sum_{K\kappa} f_{K\kappa}^{Jn} \hat{P}_{MK}^{J} | \Phi_{\kappa} \rangle ,
\end{eqnarray}
where $\kappa$ labels different qp configurations in Eq. (\ref{eq.config}), and the expansion coefficients $f_{K\kappa}^{Jn}$ can be obtained by solving the Hill-Wheeler-Griffin equation with appropriate many-body Hamiltonian, i.e., 
\begin{eqnarray} \label{eq.Hill_Wheeler}
  \sum_{K'\kappa'} \left[ \mathcal H^{J}_{K\kappa K'\kappa'}  - E^{n}_J \mathcal N^{J}_{K\kappa K'\kappa'} \right] f^{J n}_{K'\kappa'} = 0 ,
\end{eqnarray}
where the projected matrix elements for the Hamiltonian and norm are,
\begin{subequations} \label{H_and_N}
\begin{eqnarray}
  \mathcal H^{J}_{K\kappa K'\kappa'} & = & \langle \Phi_\kappa | \hat H \hat P^{J}_{KK'} | \Phi_{\kappa'} \rangle,  \\
  \mathcal N^{J}_{K\kappa K'\kappa'} & = & \langle \Phi_\kappa |        \hat P^{J}_{KK'} | \Phi_{\kappa'} \rangle. 
\end{eqnarray}
\end{subequations}
the level energies $E^n_J$ can be obtained from Eq.(\ref{eq.Hill_Wheeler}) as well.

Conventionally transition strengths are calculated in the PSM by writing the transition operators (such as Gamow-Teller transition operator for $\beta$ decay and electron-magnetic transition operators for spectroscopy) in the qp representation \cite{PSM_review, LJWang_2018_PRC_GT} instead of adopting the ROBTD. The ROBTD can be actually derived as,
\begin{align} \label{eq.new}
  & \big\langle \Psi^{n_f}_{J_f} \big\| \left[ \hat c^\dag_\mu \otimes \tilde{\hat c}_\nu \right]^\lambda \big\| \Psi^{n_i}_{J_i} \big\rangle 
  \nonumber \\
  =&\ \sqrt{2J_f+1}
      \sum_{K\kappa K'\kappa'} f^{J_f n_f}_{K \kappa} f^{J_i n_i}_{K' \kappa'} 
      \sum_\rho C^{J_f K}_{J_i K-\rho \lambda \rho} \nonumber \\
      & \quad \times
      \frac{2J_i+1}{8\pi^2} \int d\Omega D^{J_i \ast}_{K-\rho K'}(\Omega)
      \sum_{m_\mu m_\nu} C_{j_\mu m_\mu j_\nu m_\nu}^{\lambda \rho}  \nonumber \\
      & \quad \times
      (-)^{j_\nu + m_\nu}
      \big\langle \Phi_{\kappa} \big|  
      \hat c^\dag_{j_\mu m_\mu}  {\hat c}_{j_\nu -m_\nu} 
      \hat R(\Omega) \big| \Phi_{\kappa'} \big\rangle 
\end{align}
where $C$ labels the Clebsch-Gordan coefficients, and the rotated matrix elements with mixed single-particle operators can be calculated by the modern Pfaffian algorithm \cite{ZRChen_2022_PRC, Hu_2014_PLB} as,
\begin{align} \label{eq.Pf_rotated}
  &\ \big\langle \Phi_{\kappa} \big|  \ 
      \hat c^\dag_{j_\mu m_\mu}  {\hat c}_{j_\nu -m_\nu} 
      \hat R(\Omega) \big| \Phi_{\kappa'} \big\rangle \nonumber \\
  =&\ \big\langle \Phi^{(b)} \big| \ \hat{b}_1 \cdots \hat{b}_n \ 
      \hat c^\dag_{j_\mu m_\mu}  {\hat c}_{j_\nu -m_\nu} 
      \hat R (\Omega) \ \hat{a}^\dag_{1'} \cdots \hat{a}^\dag_{n'} \  
      \big| \Phi^{(a)} \big\rangle \nonumber \\
  =&\ \text{Pf} (\mathbb S)  \big\langle \Phi^{(b)} \big| \hat R (\Omega) \big| \Phi^{(a)} \big\rangle
\end{align}
where $| \Phi^{(a)} \rangle$ and $| \Phi^{(b)} \rangle$ label the qp vacua of parent and daughter nuclei, with $\hat{a}^\dag$ and $\hat{b}$ being the corresponding qp creation and annihilation operators respectively. Here Pf denotes the Pfaffian of some matrix, refer to Ref. \cite{Mizusaki_2013_PLB} for the detailed definition of Pfaffians and Ref. \cite{Pfaffian_code_CPC_2011} for a numerical code for Pfaffians. 

In Eq. (\ref{eq.Pf_rotated}) $\mathbb S$ is a $(n+2+n') \times (n+2+n')$ skew-symmetric matrix, the corresponding matrix elements $\mathbb S_{ij}$ can be obtained by,  
\begin{eqnarray} \label{sij}  
  \left\{ \begin{array}{ll} 
      \frac{ \big\langle \Phi^{(b)}\big| \hat{b}_i \hat{b}_j \hat R (\Omega) \big| \Phi^{(a)} \big\rangle }{ \big\langle \Phi^{(b)} \big| \hat R (\Omega) \big| \Phi^{(a)} \big\rangle } \equiv B^{b}_{ij}(\Omega)
      & (1 \leqslant i \leqslant n, \  1 \leqslant j \leqslant n ) \\ 
      \frac{ \big\langle \Phi^{(b)}\big| \hat{b}_i \hat c^\dag_{j_\mu m_\mu} \hat R (\Omega) \big| \Phi^{(a)} \big\rangle }{ \big\langle \Phi^{(b)} \big| \hat R (\Omega) \big| \Phi^{(a)} \big\rangle } 
      & (1 \leqslant i \leqslant n, \  j = n+1 ) \\ 
      \frac{ \big\langle \Phi^{(b)}\big| \hat{b}_i \hat c_{j_\nu -m_\nu} \hat R (\Omega) \big| \Phi^{(a)} \big\rangle }{ \big\langle \Phi^{(b)} \big| \hat R (\Omega) \big| \Phi^{(a)} \big\rangle } 
      & (1 \leqslant i \leqslant n, \  j = n+2 ) \\ 
      \frac{ \big\langle \Phi^{(b)}\big| \hat{b}_i  \hat R (\Omega) \hat a^\dag_{x} \big| \Phi^{(a)} \big\rangle }{ \big\langle \Phi^{(b)} \big| \hat R (\Omega) \big| \Phi^{(a)} \big\rangle }  \equiv C^{ba}_{ij}(\Omega)
      & (1 \leqslant i \leqslant n, \  j > n+2 )  \\ 
      \frac{ \big\langle \Phi^{(b)}\big| \hat c^\dag_{j_\mu m_\mu}  \hat c_{j_\nu -m_\nu}  \hat R (\Omega) \big| \Phi^{(a)} \big\rangle }{ \big\langle \Phi^{(b)} \big| \hat R (\Omega) \big| \Phi^{(a)} \big\rangle } 
      & (i = n + 1 , \  j = n + 2 ) \\ 
      \frac{ \big\langle \Phi^{(b)}\big| \hat c^\dag_{j_\mu m_\mu}  \hat R (\Omega) \hat a^\dag_{x} \big| \Phi^{(a)} \big\rangle }{ \big\langle \Phi^{(b)} \big| \hat R (\Omega) \big| \Phi^{(a)} \big\rangle } 
      & (i = n + 1, \  j > n + 2 ) \\ 
      \frac{ \big\langle \Phi^{(b)}\big| \hat c_{j_\nu -m_\nu}  \hat R (\Omega) \hat a^\dag_{x} \big| \Phi^{(a)} \big\rangle }{ \big\langle \Phi^{(b)} \big| \hat R (\Omega) \big| \Phi^{(a)} \big\rangle } 
      & (i = n + 2, \ j > n + 2 ) \\ 
      \frac{ \big\langle \Phi^{(b)}\big| \hat R (\Omega) \hat a^\dag_{x} \hat a^\dag_{y} \big| \Phi^{(a)} \big\rangle}{ \big\langle \Phi^{(b)} \big| \hat R (\Omega) \big| \Phi^{(a)} \big\rangle } \equiv A^{a}_{ij}(\Omega)
      & (i > n + 2, \ j > n+2  ) \\ 
  \end{array} \right. \nonumber \\
\end{eqnarray}
for $i<j$, and
\begin{eqnarray}
  \mathbb S_{ij} = - \mathbb S_{ji} \qquad \textrm{for } (i>j)
\end{eqnarray}

In Eqs. (\ref{eq.Pf_rotated}, \ref{sij}) the $\big\langle \Phi^{(b)} \big| \hat R (\Omega) \big| \Phi^{(a)} \big\rangle$, $A(\Omega)$, $B(\Omega)$, $C(\Omega)$ terms are the basic overlap and contractions \cite{PSM_review, LJWang_2018_PRC_GT} that have already been prepared in our PSM work for Gamow-Teller transitions (see Ref. \cite{LJWang_2018_PRC_GT} for details). The rest five basic contractions in Eq. (\ref{sij}) should be implemented numerically in the model. The expressions of them can be derived based on Hartree-Fock-Bogoliubov theory \cite{Ring_many_body_book}, the quantum theory of angular momentum \cite{Angular_momentum_Varsha_book_1988}, and some basic algorithms of AMP \cite{PSM_review}. 

The ROBTD as shown in Eq. (\ref{eq.new}) can be used for study of allowed transition in $\beta$ decays when restricting $\lambda=0, 1$ and the indices $\mu, \nu$ to keep the parity. Then, together with the level energies $E^n_J$ in Eq. (\ref{eq.Hill_Wheeler}), they can provide important nuclear inputs for the BSG code for high-precision allowed $\beta$ spectrum, which will be studied below. The expressions in Eqs. (\ref{eq.new}, \ref{eq.Pf_rotated}, \ref{sij}) have actually many more potential applications as well. On one hand, when restricting $\lambda=0, 1, 2$ and the indices $\mu, \nu$ to change the parity, the ROBTD serves as one of the two indispensable ingredients for description of first forbidden transitions in $\beta$ decay. The PSM for first forbidden transition has been accomplished and will be published elsewhere soon. This indicates that the ROBTD developed in this work paves the way for theoretical description of the total $\beta$ and neutrino spectra by the PSM where both allowed and forbidden transitions as well as the corresponding partial decay rates and $I(\%)$ can be calculated. On the other hand, the ROBTD make it possible to study interesting physics, such as the isovector spin monopole resonance, in a straightforward way. Besides, the expressions in Eqs. (\ref{eq.new}, \ref{eq.Pf_rotated}, \ref{sij}) can be revised straightforwardly to derive the reduced two-body transition density, which will be helpful for study of the nuclear matrix elements in neutrinoless double $\beta$ decay, the $g_A$ quenching problem by chiral two-body currents in single $\beta$ decay \cite{LJWang_current_2018_Rapid, Gabriel_quench_1996_PRC, Javier2011PRL, quenching_nature_phys_2019} etc.

\section{\label{sec:result}Results and discussions}

The level energies and ROBTD from the PSM calculations can be inputted to the BSG code to study their effects on the high-precision allowed $\beta$ spectrum shape. Here we take the transition from $^{162}$Gd to $^{162}$Tb as the first example from even-even to odd-odd nuclei, and the transition from $^{103}$Ru to $^{103}$Rh as the first example for odd-mass nuclei. These examples are adopted as one single allowed transition dominates the corresponding feeding intensity $I(\%)$ and therefore plays the most important role in the total spectrum for them. Except for the phase space factor and the shape factor $C(Z, \omega)$, the other corrections in Eq. (\ref{eq.spectrum}) are all open with their parameters keeping to be the default values in the BSG code \cite{Hayen_CPC_code_2019}. 

\begin{figure}[t]
\begin{center}
  \includegraphics[width=0.48\textwidth]{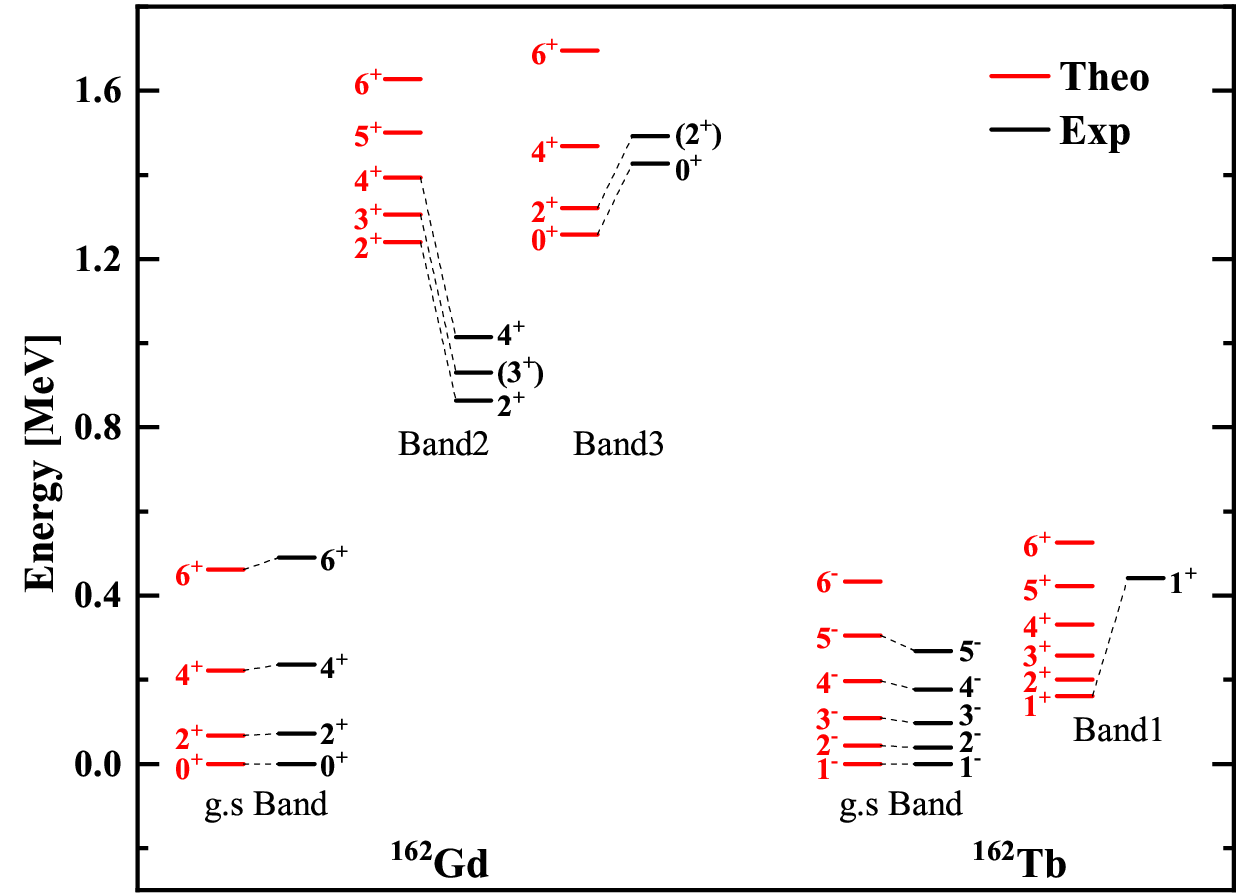}
  \caption{\label{fig:fig_2} (Color online) The calculated levels, including the ground-state (g.s.) band and other side bands, of the parent nucleus $^{162}$Gd and daughter nucleus $^{162}$Tb, as compared with available data \cite{NNDC}. } 
\end{center}
\end{figure}

Figure \ref{fig:fig_2} shows the low-lying states, including the ground-state (g.s.) band and other side bands, of parent nucleus $^{162}$Gd and daughter nucleus $^{162}$Tb calculated by the PSM as compared with the data \cite{NNDC}. It is seen that the spin and parity of g.s. for both nuclei can be reproduced well and the structures of all bands are described reasonably, while some discrepancies exist for the band-head energies of side bands. The $0^+$ g.s. of $^{162}$Gd can decay to a $1^+$ state of $^{162}$Tb by allowed transition while other relevant transitions are all forbidden. This allowed transition dominates the $\beta$-decay feeding intensity of $^{162}$Gd with $I(\%) = 95.5$, due to the strong transition strength with small Log$ft=4.46$, which is reproduced reasonably by our PSM calculations as seen from Table \ref{tab:table2}. These facts indicate that the nuclear many-body wave functions of both nuclei are described reasonably by the PSM, and the ROBTD calculated by the PSM should be appropriate. 

\begin{table}[b] 
  \caption{\label{tab:table2} The calculated Log$ft$ values for allowed transitions in the context, as compared with the corresponding data \cite{NNDC}, see text for details. }
  \begin{ruledtabular}
  \begin{tabular}{cccccc}
    \textrm{Parent} & $J_i^{\pi_i}$ & \textrm{Daughter} & $J_f^{\pi_f}$ & Log$ft_{\textrm{(Exp)}}$ & Log$ft_{\textrm{(PSM)}}$  \\ \colrule
    $^{162}$Gd      &    $0^+$      &   $^{162}$Tb      &     $1^+$     & 4.46 & 4.10  \\ 
    $^{103}$Ru      &    $3/2^+$    &   $^{103}$Rh      &     $5/2^+$   & 5.72 & 5.74  \\
  \end{tabular}
  \end{ruledtabular}
\end{table}

\begin{figure}[htbp]
\begin{center}
  \includegraphics[width=0.48\textwidth]{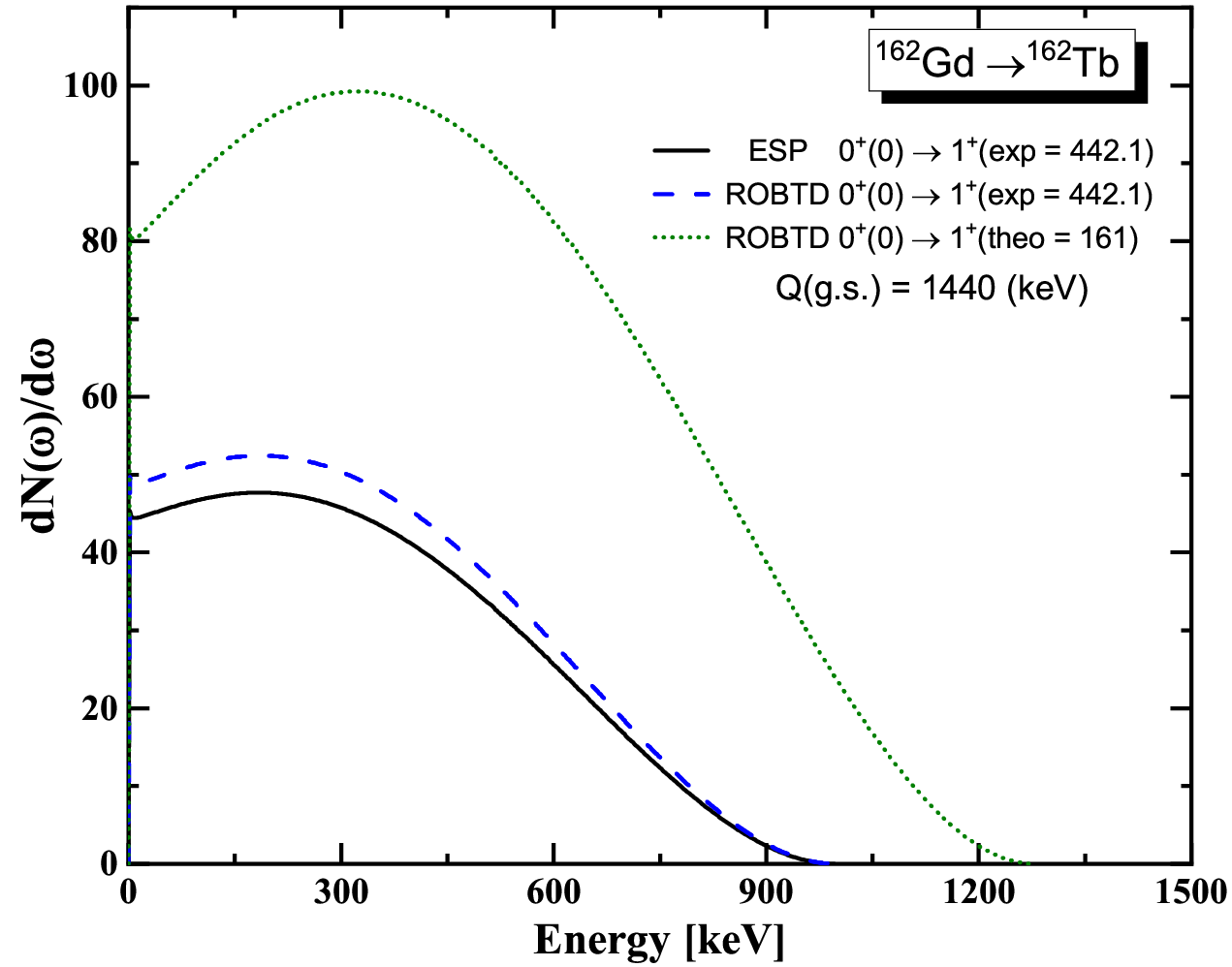}
  \caption{\label{fig:fig_3} (Color online) The $\beta$ spectrum for allowed $0^+ \rightarrow 1^+$ transition from $^{162}$Gd to $^{162}$Tb by the BSG code with different nuclear-structure inputs. See the text for details. } 
\end{center}
\end{figure}

\begin{figure}[t]
\begin{center}
  \includegraphics[width=0.48\textwidth]{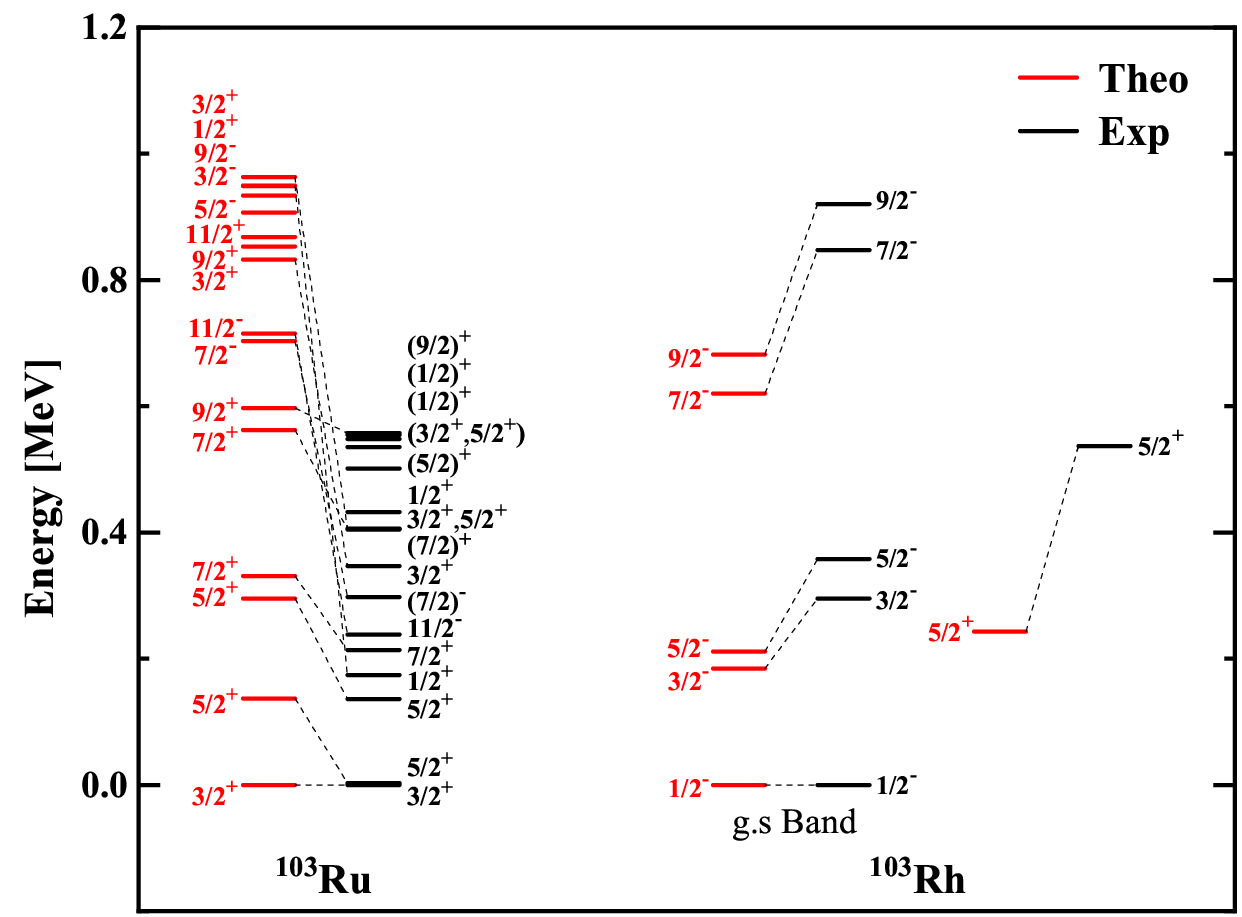}
  \caption{\label{fig:fig_4} (Color online) The same as Fig. \ref{fig:fig_2} but for the decay of $^{103}$Ru to $^{103}$Rh. } 
\end{center}
\end{figure}

\begin{figure}[htbp]
\begin{center}
  \includegraphics[width=0.48\textwidth]{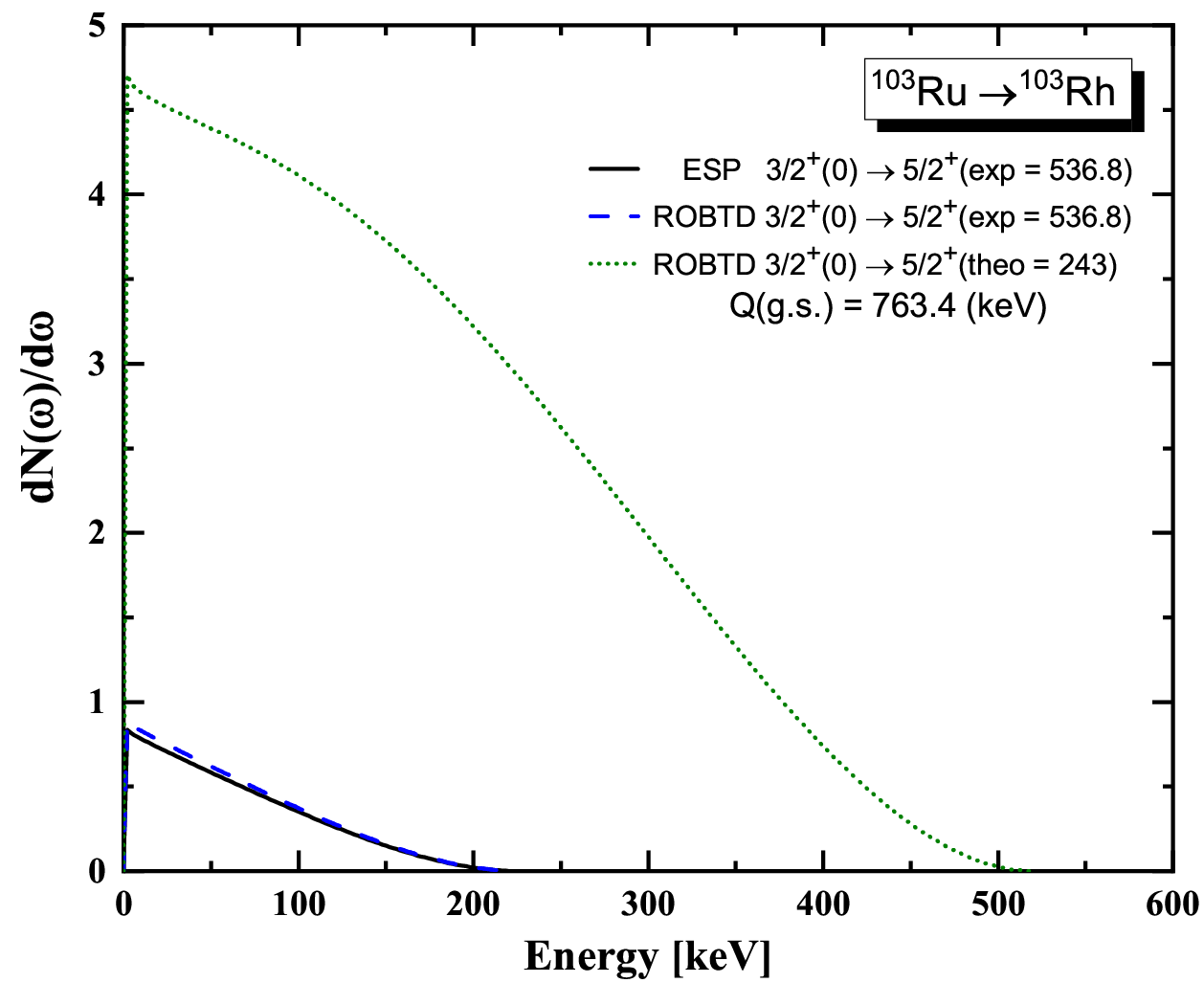}
  \caption{\label{fig:fig_5} (Color online) The same as Fig. \ref{fig:fig_3} but for the decay of $^{103}$Ru to $^{103}$Rh. See the text for details.  } 
\end{center}
\end{figure}

With the ROBTD from PSM and the experimental level energies (442.1 keV) as inputs, the high-precision $\beta$ spectrum for such a strong allowed transition is calculated by the BSG code, which is shown in Fig. \ref{fig:fig_3} (blue dashed). When no external input for ROBTD of advanced many-body method is given, the BSG code has to evaluate nuclear matrix elements for form factors in Table \ref{tab:table1} in an extreme simple single-particle (ESP) manner \cite{Hayen_CPC_code_2019}. In the ESP case, the nuclear many-body wave function is approximated by a single-particle state closest to the neutron or proton Fermi surface for the final nucleon in a filling scheme. The picture is easy to understand for odd-mass nuclei, while for even-even and odd-odd nuclei the single-particle state of the last-filling nucleon is constructed using a spectator nucleon to obtain the correct angular momentum coupling. The analytical expression of nuclear matrix elements for even-mass and odd-mass nuclei can be obtained readily in the ESP case (see the Appendix of Ref.\cite{Hayen_CPC_code_2019} for details). The ESP case corresponds to a pure single-particle manner neglecting many-nucleon correlations such as collective degrees of freedom and/or configuration mixing. Therefore the ESP evaluation may be a good approximation for very low-lying states of light and medium-heavy odd-mass nuclei, while its validity for other cases such as even-mass nuclei, heavy nuclei, transitions from isomers, transitions to excited states etc. should be tested carefully by elaborate many-body method for the nuclear matrix elements. When adopting the ESP evaluation and the experimental level energies as inputs, the $\beta$ spectrum for the allowed transition is shown in Fig. \ref{fig:fig_3} by black solid line. It is seen that many-nucleon correlations in the ROBTD of PSM are important and turn out to increase the $\beta$ spectrum by about $10\%$. Actually the $0^+$ g.s. of $^{162}$Gd is found to have collective qp vacuum as the main configuration and small mixing with a neutron 4-qp configuration, and the $1^+$ state of $^{162}$Tb has 2-qp $\nu 5/2^-[523] \otimes \pi 7/2^-[523]$ as the main configuration and about $8\%$ mixing with some 4-qp configurations. 

When the calculated level energies (161 keV) from the PSM are adopted as inputs, the corresponding $\beta$ spectrum is shown in Fig. \ref{fig:fig_3} by green dotted line. As the calculated $1^+$ state of $^{162}$Tb is lowered than the experimental level energy as shown in Fig. \ref{fig:fig_2}, the available energy for leptons of the transition, $Q_{if}$, is exaggerated, which increases the phase space factor as seen from Eqs. (\ref{eq.spectrum}, \ref{eq.Qif}) and accordingly the $\beta$ spectrum since the phase space factor and Fermi function dominate the $\beta$ spectrum shape. Although it is straightfoward to see this conclusion, one should pay attention when nuclear-structure model calculations (for ROBTD and level energies etc.) are indispensable for predicting $\beta$ and neutrino spectra of exotic nuclei with very limited experimental data to solve problems such as the reactor anti-neutrino anomaly by for example the summation method in Ref. \cite{Estienne_PRL_2019}. 

We finally report briefly the results for odd-mass nuclei. Figure \ref{fig:fig_4} represents the calculated level energies for low-lying states of parent $^{103}$Ru and daughter $^{103}$Rh nuclei. It is seen that the spin and parity of g.s. for both odd-mass nuclei are reproduced successfully. Besides, the structure of the g.s. band of $^{103}$Rh is described reasonably, including the signature-splitting behavior. The $3/2^+$ g.s. of $^{103}$Ru can decay to a $5/2^+$ state of $^{103}$Rh by allowed transition which dominates the $\beta$-decay feeding intensity with $I(\%) = 92.0$. The Log$ft$ of such a strong transition is well reproduced by the PSM calculations as seen from Table \ref{tab:table2}. The $\beta$ spectrum for such a allowed transition with different inputs, i.e., ROBTD by PSM or ESP, level energies of experimental data or PSM, is shown in Fig. \ref{fig:fig_5}. It is seen that when experimental level energies are adopted, the ROBTD from PSM tends to increase the spectrum by ESP by about $6\%$, indicating that the single-particle approximation of the ESP may be more appropriate for odd-mass nuclei. The difference in $\beta$ spectrum by the ROBTD and the ESP should come from nuclear many-nucleon correlations as well. The $3/2^+$ g.s. of $^{103}$Ru is found to have 1-qp $\nu 3/2^+[411]$ as the main configuration and $13 \%$ mixing with other 1-qp and 3-qp configurations, and the $5/2^+$ state of $^{103}$Rh has 1-qp $\pi 5/2+[422]$ as the main configuration and $38 \%$ mixing with many 1-qp and 3-qp configurations. When the calculated level energies by the PSM are adopted as inputs, the corresponding $\beta$ spectrum is shown to increase greatly due to the small $\beta$-decay $Q$ value $Q_{\beta}$(g.s.)=763.4 keV. It is worth mentioning that all the corrections in Eq. (\ref{eq.spectrum}) are expected to contribute several percentage to the $\beta$ spectrum where only purely the phase space factor and Fermi function are considered, when adopting the ESP approximation (see the Fig. 5 of Ref. \cite{Hayen_CPC_code_2019} for example).

\section{\label{sec:sum}summary and outlook}

We aim to develop a shell-model method based on angular-momentum-projection techniques, for description of nuclear total $\beta$ spectrum and neutrino spectrum, where both allowed and forbidden transitions can be treated, both large model space and configuration space are adopted for highly-excited states, the derived nuclear states are in the laboratory frame with good spin and parity, and that even-even, odd-odd and odd-mass nuclei from light to superheavy ones can be described. As the first step, we developed our projected shell model (PSM) for calculating the reduced one-body transition density (ROBTD) for nuclear $\beta$ decay (and also electron capture), by giving the analytical expression of the corresponding ROBTD with the Pfaffian algorithm. For the first application of our work, the calculated level energies and ROBTD from the PSM are inputed to the Beta Spectrum Generator (BSG) code to study the high precision $\beta$ spectrum of allowed transitions for even-mass case of $^{162}$Gd$\rightarrow$$^{162}$Tb and odd-mass case of  $^{103}$Ru$\rightarrow$$^{103}$Rh. When experimental level energies are adopted, the calculated $\beta$ spectrum by ROBTD of the PSM where nuclear many-body correlations are treated reasonably, deviates from the $\beta$ spectrum by the extreme simple-particle evaluation of the BSG by up to $10\%$. Besides, the calculated $\beta$ spectrum shows sensitive dependence on the calculated level energies from nuclear models.

The next application of our work would be the description of first forbidden transitions in $\beta$ decay, which is accomplished and will be published elsewhere soon. Then, it is possible to describe the total nuclear $\beta$ and neutrino spectra by the PSM since both allowed and forbidden transitions as well as the corresponding partial decay rates and feeding intensity can be calculated theoretically. 

Many more potential applications are also planned as near-future works, such as the isovector spin monopole resonance etc. as well as deriving the reduced two-body transition density and then studying the nuclear matrix elements in neutrinoless double $\beta$ decay, the $g_A$ quenching problem by chiral two-body currents in single $\beta$ decay etc.

\begin{acknowledgments}
  L.J.W. would like to thank X. L. Huang and J. L. Liu for motivating the studies of nuclear $\beta$ and neutrino spectra. This work is supported by the Fundamental Research Funds for the Central Universities (Grant No. SWU-KT22050), by the National Natural Science Foundation of China (Grant No. 12275225), and partially supported by the Key Laboratory of Nuclear Data (China Institute of Atomic Energy). 
\end{acknowledgments}





\bibliography{reference} 

\end{document}